\documentclass{PoS}
\usepackage{gensymb}

\title{GRANDProto300: a pathfinder with rich astroparticle and radio-astronomy science case}

\ShortTitle{GP300}

\author{\speaker{Valentin Decoene} for the GRAND collaboration\footnote{For collaboration list see PoS(ICRC2019)1177} \footnote{http://grand.cnrs.fr}\\
        Sorbonne Universit\'e, CNRS, UMR 7095, Institut d'Astrophysique de Paris, 98 bis bd Arago, 75014 Paris, France\\
        E-mail: \email{decoene@iap.fr}}

\abstract{The GRANDProto300 (GP300) experiment is the first stage of the Giant Radio Array for Neutrino Detection (GRAND) project. It will be deployed between 2020 and 2021 in a radio-quiet area, at $3000$\,m of altitude at the rim of the Tibetan plateau, over a total surface of $200$\,km$^2$. The primary goal of GP300 is to demonstrate the viability of the GRAND detection concepts. It will provide a unique test bench to develop and validate new identification and reconstruction techniques for the radio detection of very inclined air-showers, in the perspective of the next stages of GRAND. GP300 also proposes a rich science case, which includes accurate measurements of cosmic-ray and gamma-ray air-showers in the energy range of $30$\,PeV to $1$\,EeV, and a wide-field survey of the Epoch of Reionization, and of radio transients such as Giant Radio Pulses and Fast Radio Bursts. }

\FullConference{36th International Cosmic Ray Conference -ICRC2019-\\
		July 24th - August 1st, 2019\\
		Madison, WI, U.S.A.}

\begin{document}

\section{Introduction}
The GRANDProto300 (GP300) experiment is a pathfinder for the Giant Radio Array for Neutrino Detection (GRAND). GRAND will be a network of about 20 sub-arrays of approximately $10\,000$ radio antennas each, totalling a combined area of $200\,000$\,km$^2$. It will constitute a unique observatory for ultra-high-energy cosmic particles (neutrinos, cosmic rays and gamma rays) \cite{grand:olivier}.

GP300 has multiple goals. First of all, GP300 will validate the GRAND detection principle of neutrino-induced near-horizontal extensive air showers (EAS). To do so, it will have to demonstrate that air showers following very inclined trajectories --such as the ones induced by neutrinos-- can be detected using a standalone radio array with high efficiency and a rejection of background transient events close to $100\%$. With an adequate detector setup (presented in Section~\ref{detector}), GRANDProto300 will also be an ideal test bench to evaluate the quality of the reconstruction procedure (Section~\ref{performances}) for the direction, energy and nature of the primary particle, which can be achieved for very inclined EAS. 

Finally, GP300 will not only be an engineering prototype, but also an instrument to study the physics of air showers and cosmic rays, and an observatory of astrophysical phenomena (Section~\ref{science_case}).

\section{The GP300 detector} \label{detector}
The GP300 detector will be an array of $300$ detection units to be deployed between 2020 and 2021 over a $200$\,km$^2$ radio-quiet area located at an elevation of $3000$\,m, in the Qinghai province.

    \subsection{Deployment site and layout} \label{site}
    The quality of the radio environment is of paramount importance for the autonomous radio detection of air showers. A dozen of thorough measurement campaigns were carried out between August 2017 and March 2019. The two principal requests for a suitable site were that the integrated power of the stationary noise level ($\sigma$) in the $50-200$\,MHz be not larger than twice the irreducible level due to Galactic and thermal ground emission, and that the rate of transient events be below $1$\,kHz for a 6$\sigma$ threshold. Six among the nine distinct geographical sites surveyed throughout China complied with these requirements. A site on the rim of the Tibetan plateau, close to the village of Lenghu (Qinghai province) was eventually selected in July 2019. This choice was also motivated by the strong support by local authorities to the project, and the creation of a $2500$\,km$^2$ radio-protected area surrounding the site for a $10$-year period. The GRAND collaboration is presently discussing the details of the contract together with the local authorities. 
    
    Long-term measurement facilities will be deployed in summer 2019 to characterise the time dependent behaviour of the radio environment and prepare the array deployment.
    
    The baseline layout of GP300 is a square grid of $1$\,km inter-antenna spacing, deployed over a $200$\,km$^2$ area, with an infill of 85 antennas with 500\,m spacing and 27 more with a 250\,m spacing (see Fig.~\ref{fig:exposure_layout}).
    
    \subsection{Detector set up} \label{setup}
    
    \subsubsection{Antennas} \label{antenna}
    The GP300 detection unit will be composed of the {\sc HorizonAntenna} \cite{GRAND}, an active bow-tie antenna with a relatively flat response as function of azimuthal direction and frequency. Its design is inspired by the {\it butterfly antenna} developed for CODALEMA \cite{codalema} and later used in AERA \cite{aera}. It was then adapted to GRANDProto35 \cite{gp35} to allow for the measurement of the vertical polarization in addition to the two horizontal ones. The GP300 radiator size is about $2$ times smaller than for GRANDProto35 \cite{gp35} giving an optimal antenna sensitivity in the $50-200$\,MHz frequency range ($30-100$\,MHz for GRANDProto35). The motivations for this shift to higher frequency are the following: first, effects of ground diffraction scale as $\lambda/h$, where $h$ is the detector height above ground and $\lambda$ is the wavelength. To minimize this ground effect, we place the {\sc HorizonAntenna} atop a wooden pole at $h\sim5$\,m, and we operate at frequencies larger than $50$\,MHz. We then set the upper limit of the frequency range at $200$\,MHz (instead of the $80$\,MHz or $100$\,MHz used in most existing arrays) to be able to detect the Cherenkov-cone using radio. In addition, it has been shown that the signal-to-noise ratio (SNR) is higher for this frequency range \cite{balagopal} while the reconstruction also benefits from larger frequency ranges \cite{jansen, escudie, welling}. Finally a smaller antenna makes deployment easier and the mechanical structure more robust.
    
    \subsubsection{Front-end electronics}
    An electronic treatment of the antenna signal is performed at the antenna level. The analog signal is first filtered in the $30-230$\,MHz frequency band, with a difference in group delays limited to $10$\,ns. It is then digitised using a 14-bit ADC (AD9694) running at a sampling rate of $500$\,Megasamples/s. The total power consumption of the digitization process for all channels is slightly more than $2$\,W. 
    
    Digitized data is then processed inside a Zync FPGA with hardcore CPU (Xilinx XCZU5CG). The programmable logic will be used to remove narrow band sources and create a fast logic trigger. Furthermore, it will create a real-time Fourier transform of the data that can be used for environmental monitoring and searches for astronomical phenomena (see section \ref{radiotransient}). A timestamp with a precision better than $10$\,ns will be added to the triggered ADC data in order to be able to combine the data from different antennas (see section \ref{trigger}). The data is then read by the CPU where event fragments and frequency data is further analyzed, stored and transmitted to the central Central Data Acquisition (DAQ).
    The total consumption for this chip is estimated to be below $4$\,W.
    
    Communication and data transmission will use WiFi technology. The \emph{Ubiquity airMAX-AC} system fulfils our needs of reliability and transmission bandwidth (see section \ref{comms}). At the antenna side, a \emph{BULLET-AC} will be used. The maximal power consumption of these units is $8$\,W, however the average power consumption has been measured to be about $3$\,W.
    
    $100$\,W-solar panels will allow for a $100\%$ duty cycle of the unit.
    
    \subsubsection{Trigger and data throughput} \label{trigger}
    The trigger strategy is based on three trigger levels, from the antenna level, up to the whole array level.
    A first trigger level T0 is generated when the signal amplitude of one antenna channel exceeds a threshold of $5\,\sigma$, where $\sigma$ is the mean stationary noise at the antenna output. The second level T1 includes a pulse shape analysis. Previous experiments \cite{trend} have shown that a treatment based on pulse duration  allows to reject $95\%$ of the background events. For each T1, a GPS timestamp is sent to the DAQ. A $4$-byte long timestamp, and a $1$\,kHz rate at the T1 trigger level would lead to a data rate of $4$\,kB/s per detection unit. The DAQ analyzes the T1 timestamps, and issues a T2 trigger when time coincidences are found among a minimum of five detection units. Then a $3$\,$\mu$s-long time-traces from the detection units with the T2 trigger are transferred to the DAQ. At this stage, the DAQ is designed to be able to manage a $10$\,Hz rate per detection unit, but direct extrapolations from TREND results lead to an estimated T2 rate of $10$\,mHz \cite{trendICRC2015} providing a safe margin. The final data rate per detection unit is $65$\,kB/s accounting for ADC samples and header information ($2.5$\,kB per event).
    
    In parallel to this search for transient radio waves, each station will calculate the Fast Fourier Transform (FFT) over $4096$ samples in the $100-200$\,MHz frequency range (i.e. $25$\,kHz frequency resolution) and will sum them over periods of  $10$\,ms. The FFT will be coded in $1$\,byte only, after the spectrum has been corrected for the constant slope induced by the Galactic emission. This results in a rate of $410$\,kB/s data transferred for each station. In a point to multi-point configuration, this corresponds to a total throughput requirement of $123$\,MB/s to the central DAQ. This will be the main bottleneck for the communication system and drives the final setup.
    
    \subsubsection{Communication} \label{comms}
    The \emph{Ubiquity airMAX-AC} system relies on the WiFi communication of \emph{BULLET-AC} devices to a single \emph{ROCKET} base station.
    As stated before, each detection unit produces $465$\,kB/s of data at most. We require that the \emph{ROCKET} device works only at $60\%$ of its capacity to ensure an efficient working network (corresponding to a $62.5$\,MB/s data flow at $100\%$). Consequently up to $75$ detection units can be connected to the same \emph{ROCKET}  base station. Therefore the GP300 array will be divided in $5$ districts.

    The five \emph{ROCKET} stations will be installed atop the DAQ room, located at one corner of the radio array. Finally \emph{Ubiquity AirFiber} will connect the DAQ room to Lenghu (distant by some $90$\,km) and the outside ethernet, at a speed of $2$\,Gbps.
    
    \subsubsection{Particle detector} \label{particle_detector}
    Given the experience with Auger, Milagro, and HAWC, we know that water-Cherenkov detectors (aka. tanks) are very efficient and affordable detectors of muons, and we are planning on a hybrid prototype (with independent triggers) to both validate the horizontal reconstruction of radio signals, and have complementary measurements that enrich the physics potential of this stage of GRAND (see section \ref{science_case}).
    
    \subsection{GRANDProto300 as a test-bench}
    GP300 will also be used as a test-bench to develop, test and improve the techniques and the technology for the future stages of GRAND.
    
    Trigger strategies more elaborated than what will be implemented in the first stage of GP300  (see section \ref{trigger}) --in particular based on machine learning \cite{fuhrer}-- will be tested, as well as multi-point communications between antennas for instance. Other radically different strategies, such as stations composed of several detection units (following ARA and ARIANNA designs \cite{ara, arianna}) or phased station array may also be investigated.

\section{Detector performances} \label{performances}

    
    \subsection{Exposure}
\begin{figure}[htbp]
    \begin{center}
    \includegraphics[width=0.59\linewidth]{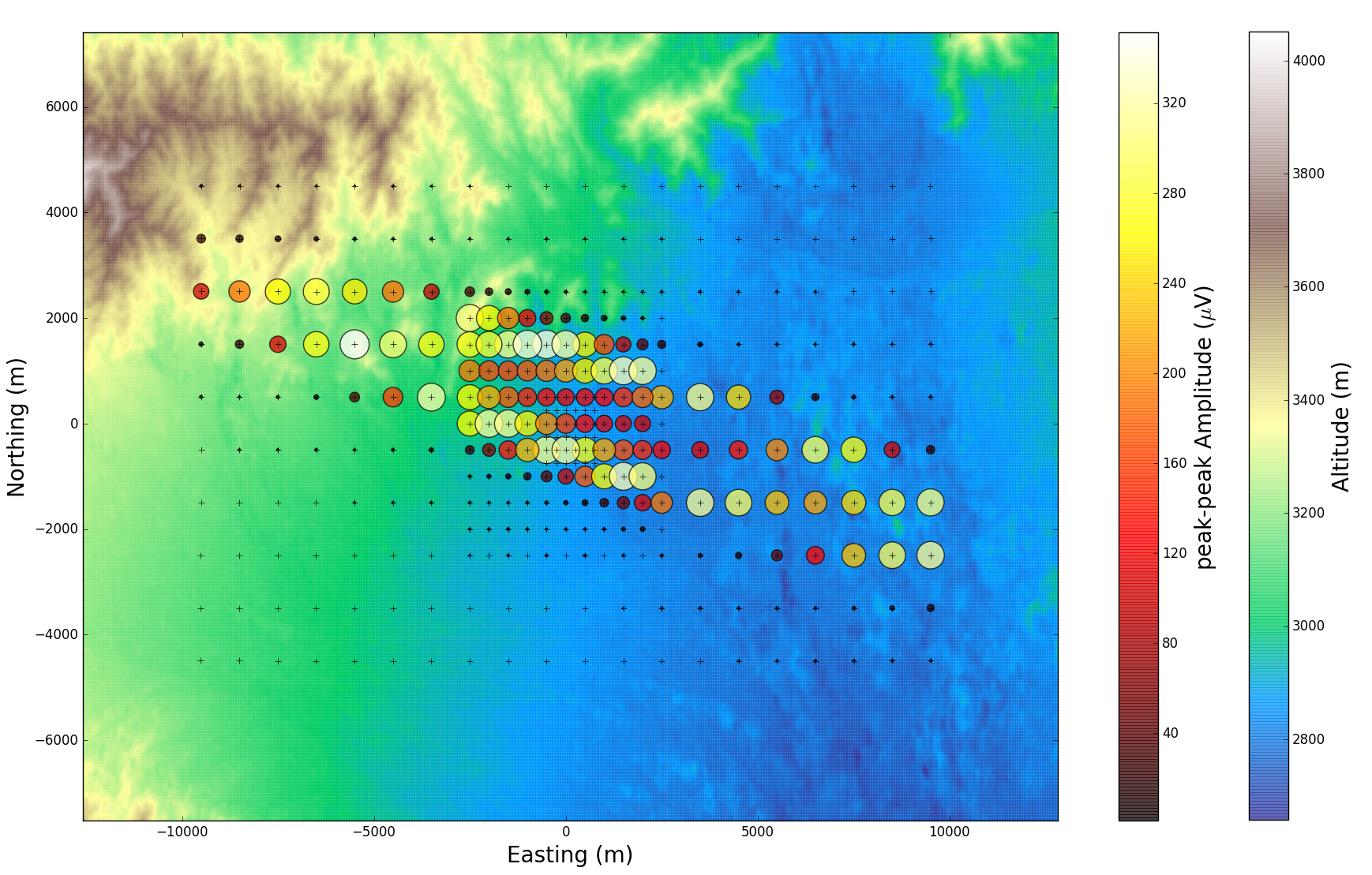}
    \includegraphics[width=0.40\linewidth]{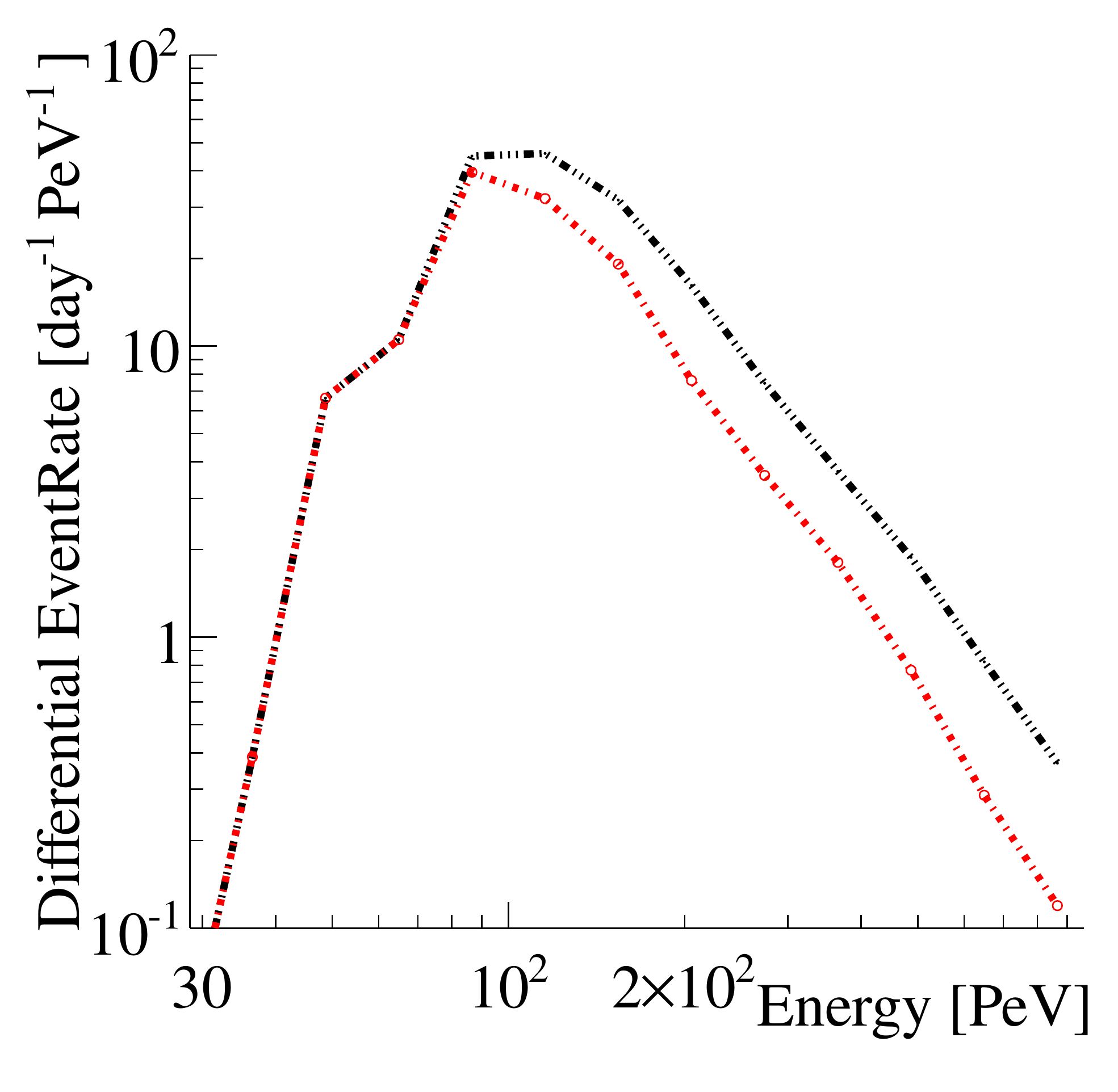}
    \caption{{\it Left:} GP300 array layout (black stars) displayed over the site topography, together with a simulated air shower, with $E=0.42$\,EeV and zenith $\theta = 83$\degree. Simulated antennas are represented by circles of radius proportional to the signal peak-to-peak amplitude at antenna output.  {\it Right:} GP300 differential event rate as a function of the primary energy for the preliminary layout. In black (small-dotted line) the total detected event rate and in red (big-dotted line) the rate of detected events with a shower core contained inside the array, which should allow for better reconstruction performances.}
    \label{fig:exposure_layout}
    \end{center}
    \end{figure}

    The GP300 exposure to cosmic rays was computed from an initial set of 5665525 trajectories sampled on a $110$ by $120$\,km area, covering all azimuths, zenith angles between $45$ and $89.9$\degree and energies between $10$ and $1000$\,PeV. A geometric pre-selection, based on a conical parametrisation of the radio-emission by the associated showers \cite{toymodel}, allowed to select $12\,000$ showers among those likely to trigger the array. These selected showers and their electromagnetic radiation were simulated using ZHAires with the Sybil2.0 hadronic model and a thinning factor of $10^{-4}$.  The shadowing effects due to the topography was also taken into account in the simulation. The antenna response was computed using the {\sc HorizonAntenna} specifications (see section \ref{antenna}) and a Butterworth numerical filtering applied on the output signal in the $50-200$\,MHz range. No noise was added in this study. Finally the trigger conditions require a minimal number of 5 antennas with a voltage signal peak-to-peak above $5\,\sigma = 75$\,$\mu$V per event, where $\sigma=15$\,$\mu$V is the mean stationnary noise expected at antenna output in the range $50-200$\,MHz\,\cite{GRAND}. One of the simulated event is displayed in Figure \ref{fig:exposure_layout}.

    The differential number of events per day (see Figure \ref{fig:exposure_layout}) was then derived from these simulations, using the spectrum obtained from the TALE measurements \cite{tale} . The infill improves the statistics at energy down to 40\,PeV, while the 200\,km$^2$ detector area makes it possible to collect significant statistics close to 1\,EeV (see Table \ref{tab:event_rate}). Additional simulations will be performed to adjust the GP300 layout and thus optimise the exposure at low energies as well as reconstruction performances. 
    
    \begin{table}[htbp]
        \centering
        \resizebox{\textwidth}{!}{%
        \begin{tabular}{lcccccccccccc}
        \hline
            E {\small [PeV]} & $30$ & $42$ & $74$ & $100$ & $133$ & $177$ & $237$ & $316$ & $562$ & $1000$ \\
            Rate [day$^{-1}$\,PeV$^{-1}$] & $ 4.1$ & $ 93.2$ & $1124.8$ & $1531.7$ & $1412.7$ & $ 957.7$ & $589.3$ & $ 391.5$ & $154.2$ & $92.3$  \\
            Contained rate & $4.1$ & $93.2$ & $ 989.7$ & $1071.3$ & $850.7$ & $451.8$ & $284.2$ & $190.5$ & $53.6$ & $29.9$ \\
        \hline
        \end{tabular}
        }
        \caption{Detected event rates per day and per energy bin. Each energy bin starts at the energy value of column $i$ and ends at the energy value of column $i+1$. The contained rates correspond to detected events with a shower core contained in the array.}
        \label{tab:event_rate}
    \end{table}
    
    \subsubsection{Reconstruction} \label{recons}
    An excellent reconstruction performance will be required for GRAND, and GP300 provides a setup to test and improve reconstruction procedures.
    
    In simulations, an angular reconstruction based on an hyperbolic wave-front shape model \cite{corstanje} reaches an angular accuracy better than $0.2$\,\degree. This result was obtained on a toy-model array configuration for a $1$-km step square grid also including  background noise effects and instrumental effects: bandpass filtering, a sampling rate of 500\,MSamples/s and a GPS jitter of more than $5$\,ns) \cite{toymodel}.
    In the case of GP300 no events below the horizon are expected to trigger the detector. Characteristics of events reconstructed below the horizon will be of major interest in order to improve the reconstruction for the next stages of GRAND. In addition a particle detector (see section \ref{particle_detector}) may also help in determining the systematic effects of the reconstruction for events above the horizon.
    
    The reconstruction of the energy and of the maximum shower elongation rate, $X_{\rm max}$, based on standard methods such as in Ref.~\cite{buitink}, might not be directly applicable in the configuration of GP300. The emission source is far from the observing array (thus limiting the resolution on $X_{\rm max}$) and the very inclined trajectories imply that the first triggered antenna corresponds to a different stage of development than the latter ones. This has a significant effect on the reconstruction, but is only marginally taken into account in standard techniques developed for vertical showers \cite{pao:recons, tunkarex:recons}.
    
    Finally the standalone radio-detection requirement of the experiment leads to less information on the event for reconstruction. The new techniques need to be more complex in order to tackle the specific geometry of inclined air showers but also more robust as it should work on radio only. These new methods are presently being developed within the GRAND collaboration.
    
\section{GP300 science case} \label{science_case}
The GP300 experiment will not only test the radio-techniques and provide a benchmark for GRAND but its design and its expected performances make it suitable for a rich science case.
    \subsection{Air-shower physics}
    For inclined air showers, the only particles reaching the ground are muons \cite{zas}. With the complementary particle detector array, a separate measurement of the shower components will be made \cite{holt}. A comparison of the muon component, measured with the particle array, and the electromagnetic (EM) energy in the shower, measured with radio, will shed light on the enhanced muon content observed in ultra-high-energy air showers \cite{aab1,aab2}. In general, the hadronic processes in air showers can be well studied when both EM and muonic information is available \cite{enterria}. 
    
    \subsection{Ultra high energy gamma rays}
    Air showers produced by UHE gamma rays with a zenith angle larger than $65$\,\degree have a dominant EM component that is fully absorbed by the atmosphere before reaching the ground (unlike cosmic ray air showers in which muons are expected to reach the ground). The particle detector will be used as a veto in order to discriminate between showers induced by UHE gamma rays and UHECR. Preliminary simulations have shown that GP300 is able to reach an identification efficiency of almost $100\%$ for zenith angles between $65$ and $85$\,\degree and energies above $10^9$\,GeV. If no gamma rays are identified among a sample of $10\,000$ showers detected in $2$ years, the limit on the fraction of UHE gamma rays will be $0.03\%$ at $95\%$\,C.L. instead of $0.1\%$ for the current best limit \cite{niechciol}.
    
    \subsection{Galactic/extra-galactic transition}
    In the energy range between $10^8$ and $10^9$\,GeV, a transition between galactic and extra-galactic cosmic ray sources is expected~\cite{dawson}. Accurate measurements of the energy, composition and distribution of the arrival directions with the large event statistics of GP300 will allow to distinguish between the different astrophysical source models.
    
    \subsection{Radio Astronomy} \label{radiotransient}
    The sum of incoherent Fourier transforms from all the antennas enables the detection of transient radio-signals. The SNR is proportional to ${N_{\rm ant}}^{1/2}$ where $N_{\rm ant}$ is the number of antennas. This technique does not reach the sensitivity of an equivalent phased array but the large number of antennas in GP300 makes it a competitive instrument. Also the whole sky can be monitored since the FoV of the array is equal to the FoV of one single antenna with a duty cycle of $100\%$. This full-sky survey would be effective for events with energies above $750$\,Jy such as Giant Pulses \cite{soglasnov, eftekhari, lorimer} and Fast Radio Bursts \cite{masui}. 
    
    The $21$-cm line from the epoch of reionization (EoR) is imprinted onto the Cosmic Microwave Background (CMB) as a line-like absorption feature, redshifted today to frequencies between $10-200$\,MHz. By measuring the temperature of the sky with mK precision, as a function of frequency, the global EoR signature can be identified as well as the absorption feature due to reionisation below $100$\,MHz. Recent experiment has found a $500$\,mK-deep absorption feature at $78$\,MHz, differing from the theoretical predictions \cite{edges}. If antennas can be calibrated with sufficient precision, GP300 could achieve $1$\,mK sensitivity using only $30$\,antennas at $80$\,MHz thus will improve the determination of the characteristics of this feature substantially.

\vspace{0.5cm}
\noindent\footnotesize{{\it Acknowledgements.} The GRAND and GRANDProto300 projects are supported by the Natural Science Fundation of China (Nos.11135010, 11375209), the Chinese Ministery of Science and Technology, the APACHE grant (ANR-16-CE31-0001) of the French Agence Nationale de la Recherche, and the France China Particle Physics
Laboratory.}

\end{document}